\documentclass[useAMS,usenatbib]{mn2e}
\usepackage{graphicx}

\def\lsim{\,\lower2truept\hbox{${<\atop\hbox{\raise4truept\hbox{$\sim$}}}$}\,}
\def\gsim{\,\lower2truept\hbox{${> \atop\hbox{\raise4truept\hbox{$\sim$}}}$}\,}

\title[HST/WFC3 deep imaging of H-ATLAS strongly lensed galaxies]
  {{\it Herschel}\thanks{{\it Herschel} is an ESA space observatory
      with science instruments provided by European-led Principal
      Investigator consortia and with important participation from
      NASA.}-ATLAS: deep HST/WFC3 imaging of strongly lensed
    submillimeter galaxies}
\author[M. Negrello et al.]
{M.~Negrello$^{1}$\thanks{mattia.negrello@oapd.inaf.it},
R.~Hopwood$^{2}$,
S.~Dye$^{3}$,
E.~da Cunha,$^{4}$,
S.~Serjeant$^{5}$, 
S.~Fleuren$^{6}$,
\newauthor 
R.~S.~Bussmann$^{7,8}$, 
A.~Cooray$^{9}$,
H.~Dannerbauer$^{10}$,
J.~Gonzalez-Nuevo$^{11}$, 
A.~Lapi$^{12,13}$,
\newauthor 
A.~Omont$^{14,15}$,
S.~Amber$^{5}$,
R.~Auld$^{15}$,
M.~Baes$^{16}$,
S.~Buttiglione$^{1}$,
A.~Cava$^{17}$,
L.~Danese$^{13}$, 
\newauthor 
A.~Dariush$^{15}$,
G.~De Zotti$^{1,13}$,
L.~Dunne$^{18}$,
S.~Eales$^{15}$,
J.~Fritz$^{16}$,
E.~Ibar$^{19}$,
R.~Ivison$^{20,21}$,
\newauthor 
S.~Kim$^{9}$, 
S.~Maddox$^{18}$,
M.~J.~Micha{\l}owski$^{21}$, 
E.~Pascale$^{15}$,
M.~Pohlen$^{15}$,
E.~Rigby$^{22}$,
\newauthor 
K. Rowlands$^{4,23}$, 
D.~J.~B.~Smith$^{24}$,
W.~Sutherland$^{6}$,
P.~Temi$^{25}$, 
J.~Wardlow$^{9}$ \\
$^{1}$INAF, Osservatorio Astronomico di Padova, Vicolo Osservatorio 5,
I-35122 Padova, Italy \\
$^{2}$Imperial College London, Blackett Laboratory, Prince Consort Road, London SW7 2AZ, UK \\
$^{3}$School of Physics and Astronomy, University of Nottingham,
University Park, Nottingham NG7 2RD, UK \\
$^{4}$Max Planck Institute for Astronomy, Koenigstuhl 17, 69117
Heidelberg, Germany \\
$^{5}$Department of Physical Sciences, The Open University, Walton
Hall, Milton Keynes MK7 6AA, UK \\ 
$^{6}$School of Mathematical Sciences, Queen Mary, University of
London, Mile End Road, London, E1 4NS, UK \\ 
$^{7}$Harvard-Smithsonian Center for Astrophysics, 60 Garden Street,
Cambridge, MA 02138, USA \\
$^{8}$Department of Astronomy, Space Science Building, Cornell
University, Ithaca, NY, 14853-6801 \\
$^{9}$Department of Physics \& Astronomy, University of California,
Irvine, CA 92697, USA \\
$^{10}$Institut fur Astronomie, Universitat Wien, Turkenschanzstraße
17, A-1160 Wien, Austria \\
$^{11}$Instituto de Fisica de Cantabria (CSIC-UC), Avda. los Castros s/n, 39005 Santander, Spain \\
$^{12}$Dipartimento di Fisica, Universita' “Tor Vergata”, Via della
Ricerca Scientifica 1, 00133 Roma, Italy \\
$^{13}$Astrophysics Sector, SISSA, Via Bonomea 265, 34136 Trieste,
Italy \\
$^{14}$UPMC Univ. Paris 06, UMR7095, Institut d'Astrophysique de
Paris, 75014 Paris, France \\
$^{15}$CNRS, UMR7095, Institut d'Astrophysique de Paris, 75014 Paris,
France \\
$^{15}$School of Physics and Astronomy, Cardiff University, The
Parade, Cardiff CF24 3AA, UK \\
$^{16}$Sterrenkundig Observatorium, Universiteit Gent, Krijgslaan 281 S9, B-9000 Gent, Belgium \\
$^{17}$Observatoire de Gen{\`e}ve, Universit{\'e} de Gen{\`e}ve, 51 Ch. des
Maillettes, 1290 Versoix, Switzerland \\
$^{18}$Department of Physics and Astronomy, University of Canterbury,
Private Bag 4800, Christchurch, New Zealand \\
$^{19}$ Departamento de Astronomıa y Astrofısica, Universidad Catolica de Chile, Vicuna Mackenna 4860, Casilla 306, Santiago 22, Chile \\
$^{20}$UK Astronomy Technology Center, Royal Observatory Edinburgh, Edinburgh EH9 3HJ, UK \\
$^{21}$Scottish Universities Physics Alliance, Institute for Astronomy, University of Edinburgh, Royal Observatory, Edinburgh EH9 3HJ, UK \\
$^{22}$Leiden Observatory, P.O. Box 9513, 2300 RA, Leiden, The
Netherlands \\
$^{23}$(SUPA) School of Physics \& Astronomy, University of St
Andrews, North Haugh, St Andrews, KY16 9SS, UK \\
$^{24}$Centre for Astrophysics Research, Science $\&$ Technology
Research Institute, University of Hertfordshire, Herts AL10 9AB, UK \\
$^{25}$Astrophysics Branch, NASA/Ames Research Center, MS 245-6, Moffett Field, CA 94035, USA 
}

\pagerange{\pageref{firstpage}--\pageref{lastpage}} \pubyear{2014}

\def\LaTeX{L\kern-.36em\raise.3ex\hbox{a}\kern-.15em
    T\kern-.1667em\lower.7ex\hbox{E}\kern-.125emX}

\begin{document}

\label{firstpage}

\maketitle

\begin{abstract}
We report on deep near-infrared observations obtained with the Wide
Field Camera 3 (WFC3) onboard the {\it Hubble} Space Telescope ({\it
  HST}) of the first five confirmed gravitational lensing events
discovered by the {\it Herschel} Astrophysical Terahertz Large Area
Survey (H-ATLAS). 
We succeed in disentangling the
background galaxy from the lens to gain separate photometry of the two
components. The HST data allow us to significantly improve on previous
constraints of the mass in stars of the lensed galaxy and to
perform accurate lens modelling of these systems, as described in the accompanying paper by Dye et al. (2013). We fit the
spectral energy distributions of the background sources from near-IR
to millimetre wavelengths and use the magnification factors estimated
by Dye et al. to derive the {\it
  intrinsic} properties of the lensed galaxies. We find these
galaxies to have star-formations rates
SFR$\sim400-2000$\,M$_{\odot}$\,yr$^{-1}$, with
$\sim(6-25)\times$10$^{10}$\,M$_{\odot}$ of their baryonic mass
 already turned into stars. At these rates of star formation, all 
 remaining molecular gas will be exhausted in less than $\sim$100\,Myr,
reaching a final mass in stars of a few $10^{11}\,$M$_{\odot}$. These galaxies are thus
proto-ellipticals caught during their major episode of star formation,
and observed at the peak epoch ($z\sim1.5-3$) of the cosmic star formation
history of the Universe.
\end{abstract}


\section{Introduction}

Recent evidence has found that almost all of high-redshift ($z\gsim1$)
dust-obscured star forming galaxies selected in the sub-millimetre (hereafter sub-mm galaxies, or SMGs) with flux density above $\sim$100$\,$mJy at 500$\,\mu$m are gravitationally lensed by a foreground galaxy or a group/cluster of galaxies \citep{Neg10,Con11,Cox11,Buss13,Fu12,Wardlow13}. These sub-mm bright sources are rare, their surface density being $\lsim$0.3$\,$deg$^{-2}$ at F$_{\rm 500}\gsim100\,$mJy \citep{Neg07} and therefore only detectable in wide-area sub-mm surveys. In fact, sub-mm surveys before the
advent of the {\it Herschel} Space Observatory \citep{Pil10} were
either limited to small areas of the sky (i.e. $<1\,$deg$^{2}$) or
severely affected by source confusion due to poor spatial
resolution \cite[e.g.][]{Coppin06,Devlin09,Weib09}. \\

The {\it Herschel} Astrophysical Terahertz Large Area Survey\footnote{www.h-atlas.org}  (H-ATLAS)
\citep{E10} is the widest area extragalactic survey undertaken with
{\it Herschel}. It has mapped $\sim570\,$deg$^{2}$ in five bands from
100 to $500\,\mu$m, down to around the $250-500\,\mu$m confusion
limit. The first 16\,deg$^{2}$ were observed during the Science
Demonstration Phase (SDP) and detected 10 extragalactic sources with $F_{\rm
  500}\ge100\,$mJy. 
Existing shallow optical and radio data clearly identifies four of these as low redshift (i.e. $z<0.1$) spiral galaxies \citep{Baes10}  and one as a radio bright  ($F_{\rm 1.4GHz}>100\,$mJy) blazar at $z=1$ \citep{Gonz10}, while the remaining five have sub-mm colours (i.e. 250$\mu$m/350$\mu$m vs 350$\mu$m/500$\mu$m flux ratios) indicative of dusty star-forming galaxies at $z>1$.
If SMGs have a steep luminosity function, as several models suggest
\citep{Gra04, Lapi06} and recent results support
\citep{E10b,Lapi11,Gruppioni13}, their number counts are expected to
exhibit an extremely sharp cut-off at bright fluxes
($\simeq80-100$\,mJy at $500\,\mu$m). This cut-off implies that only
SMGs that have had their flux boosted by an event of gravitational
lensing can be detected above this brightness threshold
\citet[][see also Fig.\,1 of Negrello et al. 2010]{Neg07}.. \\

To confirm this prediction, the five sources with $F_{\rm
  500}>100\,$mJy and with high$-z$ colours identified in the
H-ATLAS/SDP field have been the subject of intensive multiwavelength
follow-up observations.  The follow-up campaign includes observations
from the ground with the Keck telescopes \citep[][N10 hereafter]{Neg10}, the
Submillimeter Array (SMA) \citep{Neg10,Buss13}, the Zpectrometer
instrument on the NRAO Robert C. Byrd Green Bank Telescope (GBT)
\citep{Fra11,Harris12}, the Z-Spec spectrometer \citep{Lupu12}, the
IRAM Plateau de Bure Interferometer (PdBI) \citep{Om11,Om13,George14}, the
Max-Planck Millimeter Bolometer (MAMBO) at the IRAM 30 meter telescope
on Pico Veleta (Dannerbauer et al. in prep.), the Combined Array for
Research in Millimeter-wave Astronomy (CARMA) (Leew et al. in
prep.), the Jansky Very Large Array (JVLA; Ivison et al. in prep.) and also from space with the {\it Spitzer} Space Telescope
\citep{Hop11} and the {\it Herschel}/SPIRE Fourier Transform
Spectrometer \citep{Val11}. The detection, in these objects, of carbon
monoxide (CO) rotational line emission, which is a tracer of molecular
gas associated to star forming environments, has provided redshifts in
the range $z\sim$1.5-3, consistent with what can be inferred from
their sub-mm colours (N10). In contrast, the
same sources are reliably associated with lower redshift ($z<1$)
galaxies detected in the Sloan Digital Sky Survey (SDSS)
\citep{Smith11} and in the VISTA Kilo-degree INfrared Galaxy (VIKING)
Survey \citep{Fleuren12}, thus confirming the presence of a foreground
galaxy acting as a lens. In four of these systems the background
galaxy has been clearly resolved into multiple images at 880$\mu$m
with the SMA \citep[N10,][]{Buss13} thus providing the definitive confirmation of the lensing hypothesis. As part of this extensive follow-up campaign weobtained observations
in the near-IR with the Wide Field Camera-3 (WFC3) onboard the {\it
  Hubble} Space Telescope (HST) during cycle-18, using the wide-J
filter F110W and the wide-H filter F160W. \\

In this paper, we report on the results of these observations. We
exploit the sub-arcsecond spatial resolution and sensitivity of the
HST observations to disentangle the background source from the
foreground galaxy to constrain the near-IR emission of the two
components separately. A detailed lens modelling of these systems using
a ``semilinear inversion approach'' is presented in an accompanying
paper \citep[][D13 hereafter]{Dye13}.
The work is organised as follows. In Sec.\,2 we present the HST
data. In Sec.\,3 we discuss other ancillary data used to build the panchromatic
spectral energy distribution (SED) of the sources. The subtraction of
the foreground lens and the measurement of the photometry of the
the lens and the background galaxy are discussed in Sec\,4. A fit to the SED of the
lensed galaxy, from optical to millimetre wavelength, with the
addition of the  near-IR HST points, is performed in Sec.\,5. The
results are discussed in Sec.\,6 while Sec.\,7 summarises the main conclusions.

\begin{table}
 \caption{Total exposure times for observations taken with HST/WFC3 using the F110W and F160W filters.}
 \label{tab:exp_times}
 \centering
 \begin{tabular}{@{}lrr}
  \hline
  \hline
  H-ATLAS ID  &  F110W  &  F160W \\
              &  (sec)  &  (sec) \\ 
  \hline
  SDP.9   & 1412 & 3718 \\
  SDP.11  & 1412 & 3718 \\
  SDP.17  & 1412 & 3718 \\
  SDP.81  & 712  & 4418 \\
  SDP.130 & 712  & 4418 \\
  \hline
  \hline
 \end{tabular}
\end{table}

\begin{figure*}
  \hspace{-3.0cm}
  \begin{minipage}[b]{1.0\linewidth}
    \centering \resizebox{1.1\hsize}{!}{
      \hspace{+2.0cm}\includegraphics{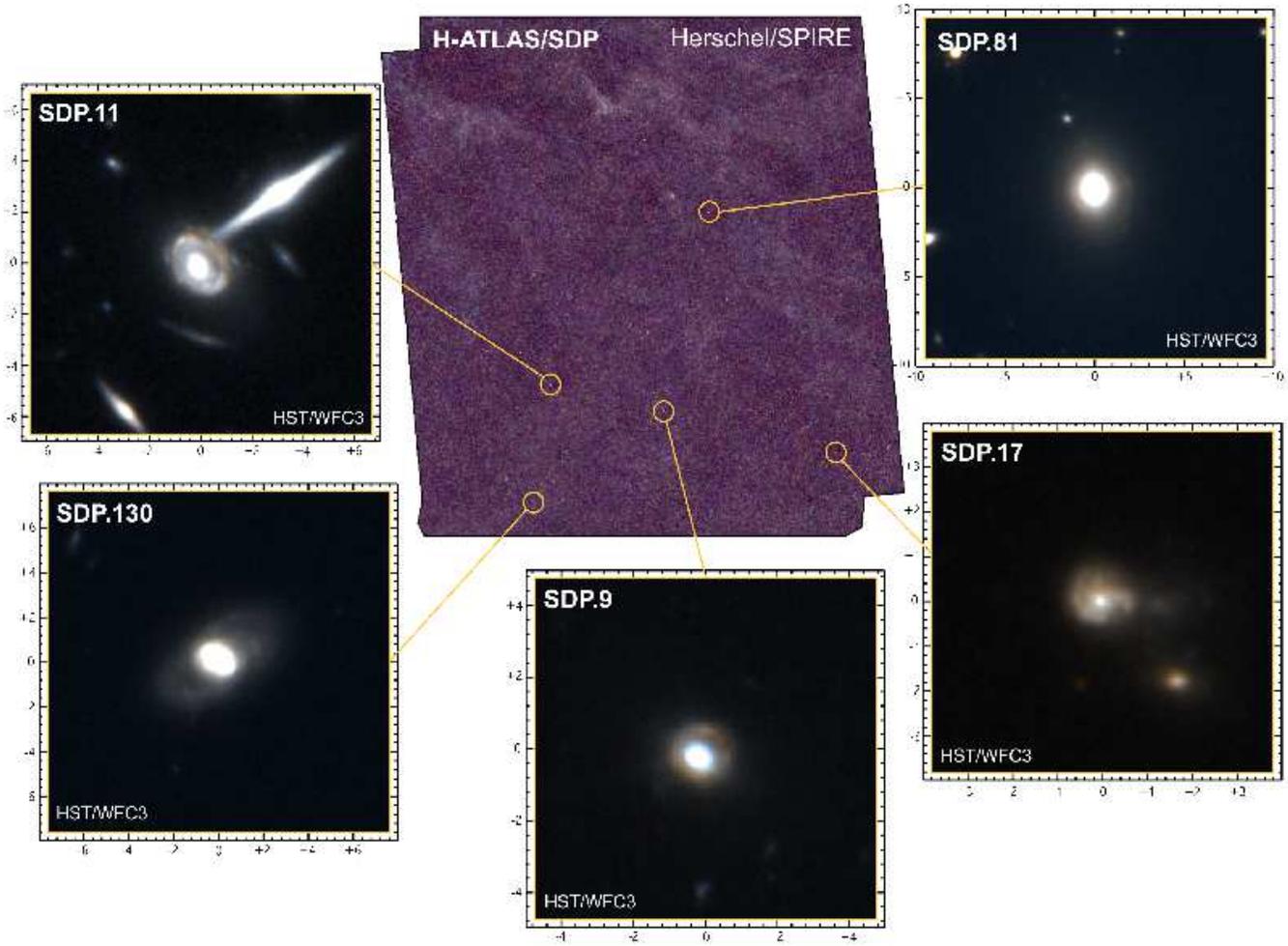}} 
  \end{minipage}
 \vspace{0.cm}
 \caption{Two-colour postage stamp HST/WFC3 images of the first five confirmed
   gravitational lensing systems discovered by H-ATLAS (blue for F110W
   and red for F160W). The position
   of the five sources in the {\it
     Herschel}/SPIRE map of the H-ATLAS SDP field is indicated by the
   yellow circles. The scale of the postage stamps is given in arcseconds.}
 \label{fig:HST_images1}
\end{figure*}

\begin{figure*}
  \hspace{-3.0cm}
  \begin{minipage}[b]{1.0\linewidth}
    \centering \resizebox{1.55\hsize}{!}{
      \hspace{-3.0cm}\includegraphics{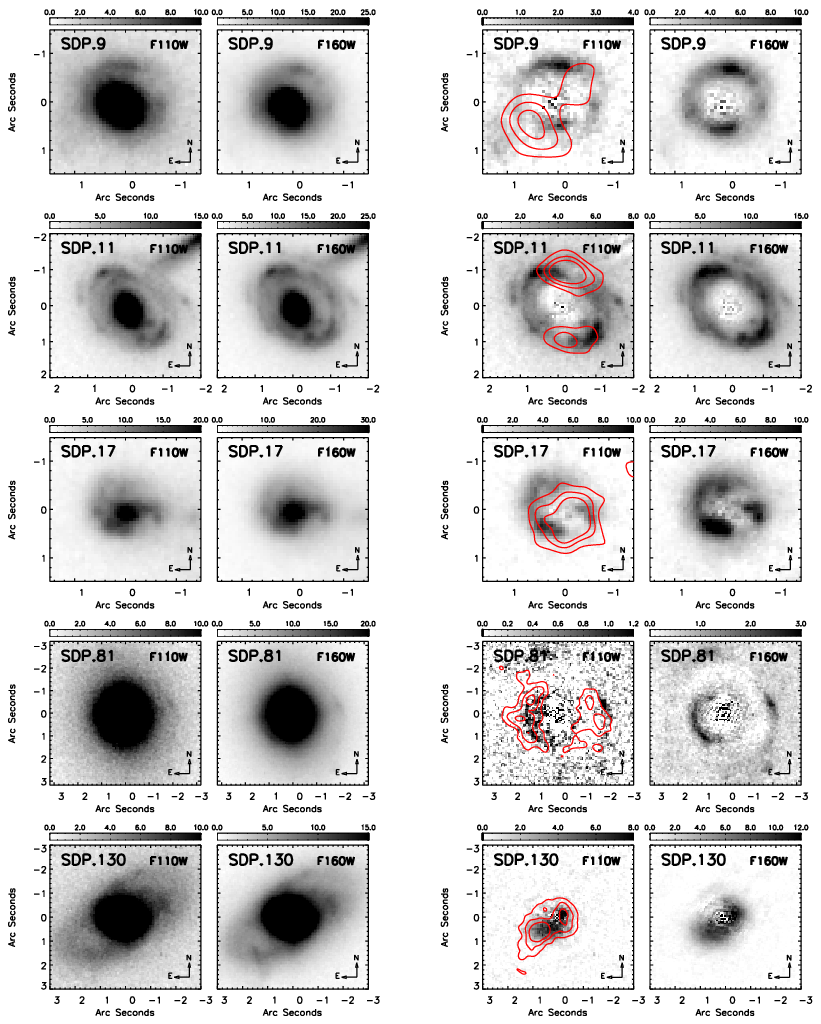}} 
  \end{minipage}
 \vspace{+0.6cm}
 \caption{HST/WFC3 images taken with the F110W and the F160W filters
   (left panels) of the first gravitational lensing events discovered by
   H-ATLAS \citep{Neg10}. The corresponding lens subtracted images
   are shown in the right panels. The colour code represents the surface
   brightness in $\mu$Jy/arcsec$^{2}$. Signal-to-noise ratio contours at
   880\,$\mu$m from the SMA (Bussmann et al. 2013) are shown
   against the lens subtrated F110W images (red curves, in steps of 3,
   6 and 9).}
 \label{fig:HST_images2}
\end{figure*}

\section{HST data}

HST observations of the five lens candidates presented in N10 were taken in April 2011 as part of the cycle-18 proposal 12194 (PI: Negrello) using 10 orbits in total, 2 for each target. Observations were made with the WFC3 using the wide-J filter F110W (peak wavelength 1.15\,$\mu$m) and the wide-H filter F160W (peak wavelength $1.545\,\mu$m), in order to maximise the chance of detection of the background galaxy, whose emission at shorter wavelengths is expected to be dominated by the foreground galaxy. About one and a half orbits were dedicated to observations in the H-band with only half an orbit (or less) spent for observations with the F110W filter. This relatively short exposure was aimed at revealing the morphology of the lens, with minimal contamination from the background source. 
The total exposure times are reported in Table\,\ref{tab:exp_times}. 
Data were reduced using the IRAF MultiDrizzle package. The pixel scale
of the Infrared-Camera is 0.128$^{\prime\prime}$ but we resampled the images to a finer pixel
scale of 0.064$^{\prime\prime}$ by exploiting the adopted
dither strategy (a sub-pixel dither patter). This provides us with a
better sampling of the point spread function whose full width at half
maximum (FWHM) is $\sim0.13^{\prime\prime}-0.16^{\prime\prime}$ at
wavelengths $\lambda=1.1-1.6\,\mu$m. Cosmic ray rejections and alignments
of the individual frames were also addressed before combining and
rebinning the images. 
Multidrizzle parameters were optimised to the final image quality.
HST cut-outs around the five targets are shown in
Fig.\,\ref{fig:HST_images1} and in the left panels of Fig.\,\ref{fig:HST_images2}. 
Due to the relatively longer integration times, the combined F160W
images exhibit higher signal-to-noise ratio to those obtained with the
F110W filter; however the main features revealed in the H-band are
also captured with the shorter exposures in the J-band\footnote{A
  cycle-19 HST/WFC3/F110W snapshot program has provided imaging data for $\gsim$100 lens candidates identified in {\it H}-ATLAS (PID: 12488; PI: Negrello).}.

\begin{figure*}
\vspace{0.0cm}
\hspace{-0.5cm}
\makebox[\textwidth][c]{
\includegraphics[width=1.\textwidth]{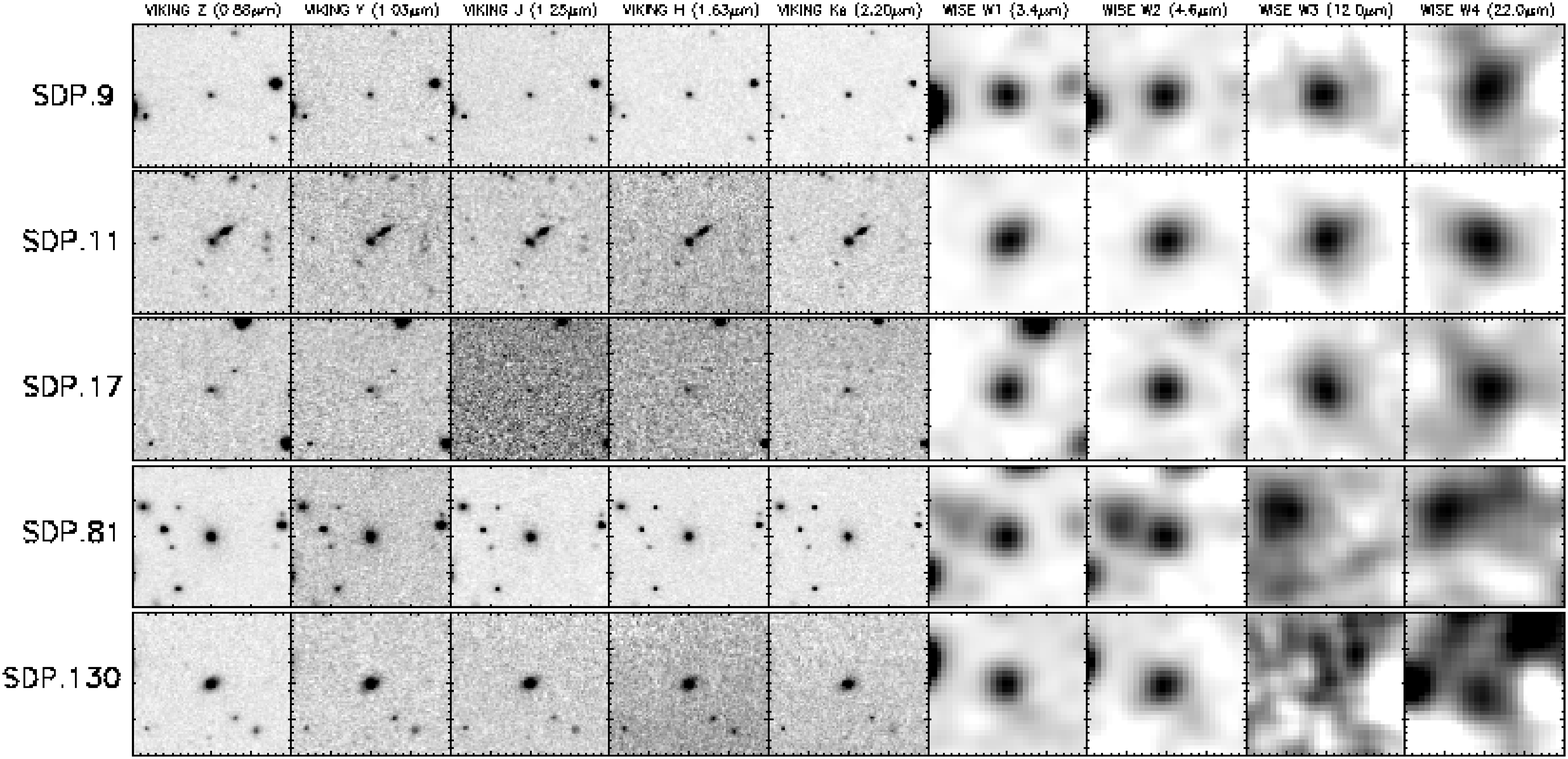}
} 
\vspace{-0.2cm}
\caption{40$^{\prime\prime}\times$40$^{\prime\prime}$ postage
  stamp images of the five
  H-ATLAS/SDP lensing systems at near- to mid-infrared wavelengths obtained 
  from the VIKING and WISE surveys. The stamps are centred at the position of the
  lensed galaxy.}
\label{fig:vik_wise}
\end{figure*}

\section{Ancillary data}

These HST  images represent the latest addition to the already
substantial set of photometric and spectroscopic data for these
sources that are reported in Table\,\ref{tab:phot_param}, and briefly summarised below. 

\subsection{Far-infrared and sub-mm/mm}

Flux density estimates at 100 to 500$\,\mu$m are provided by {\it
  Herschel}/PACS \citep{Pog10} and {\it Herschel}/SPIRE \citep{Gri10},
which are used in parallel mode for {\it H}-ATLAS. A description of
the map-making for the PACS and SPIRE data of the {\it H}-ATLAS/SDP
field can be found in \cite{Ibar10} and \cite{Pa10}, respectively,
while details of the source extraction and flux measurements are given
in \cite{Rig11}. The achieved 5$\,\sigma$ detection limits (including
source confusion), are 33.5 to 44.0 mJy/beam from 250 to 500\,$\mu$m,
132\,mJy/beam at 100\,$\mu$m and 121$\,$mJy/beam at 160$\,\mu$m. The
five sources discussed here have, by selection, a flux density above
100\,mJy at 500\,$\mu$m (see Table\,\ref{tab:phot_param}) and are
therefore robustly detected at the SPIRE wavebands. However only 3 of
them are detected in PACS at more than 3$\,\sigma$, namely SDP.9, SDP.11
and SDP.17. One source, SDP.81, was undetected while the other,
SDP.130, falls outside the region covered by PACS in parallel
mode. 
Deeper PACS minimaps of these two objects at 70$\mu$m and 160$\mu$m
were obtained by \cite{Val11}. Both source were detected at
160$\mu$m while upper limits on their flux density were obtained at 70\,$\mu$m. \\
Follow-up observations with the SMA \citep[N10,][]{Buss13} and
IRAM/MAMBO (N10, Dunnerbauer et al. in prep.) provide flux estimates
for all five targets at 880\,$\mu$m and at 1200\,$\mu$m,
respectively. 

\subsection{Optical}

The {\it H}-ATLAS/SDP field is covered by the
Sloan Digital Sky Survey. Four of the H-ATLAS/SDP lenses have a
reliable association in SDSS with $r<22.40$ \citep{Smith11}, the
exception being SDP.11, whose optical counterpart has
$r=22.41$. The SDSS flux densities used for the SED fitting in Sec.\,5 are those derived from the Data Release 7 model magnitudes
(see also N10).  \\
Dedicated follow-up observations with the Keck telescope provided
supplementary optical imaging in the $g$ and $i$ bands. As discussed
in N10, the lensed source is undetected in the optical. The optical
flux densities reported in  Tab.\,\ref{tab:phot_param}, derived from
the light-profile modeling as described in N10, refer to either
the whole system (lens+source) or the lens alone when the latter is
completely dominating over the background galaxy as suggested by the HST
imaging data. Upper limits on the optical emission from the background
source are also shown in the table. These limits were derived after subtracting
the best-fit model for the light profile. The local standard deviation was
scaled to the area of a ring of radius 1.5$^{\prime\prime}$ (inner radius 
of 1$^{\prime\prime}$ and outer radius of 2$^{\prime\prime}$). The
limits are not reported for SDP.17; in fact, in this case the HST data
suggest (Sec.\,5) that at optical wavelengths the emission of the
source might not be negligible, implying that the GALFIT model derived from the
Keck image carries contributions from both the lens and the background galaxy.

\subsection{Near- and mid-infrared}

Near-IR imaging data are available through the UKIRT Infrared Deep Sky
Survey (UKIDSS), Large Area Survey (LAS) and the VISTA Kilo-degree
INfrared Galaxy (VIKING; Sutherland et al., in prep.) survey
\citep[see also][]{Fleuren12}. The VIKING survey is 1.4 magnitudes
deeper than UKIDSS/LAS, so we use VIKING data only in the
present work. The VIKING survey provides photometric measurements in 5 broad-band filters: $Z$, $Y$, $ J$, $H$, and $Ks$, down to a typical 5$\,\sigma$ magnitude
limit of 21.0 in J-band and 19.2 in Ks-band (in the Vega system). The
median image quality is $\sim$0.9$^{\prime\prime}$. All our targets
are found to have a reliable association in the 
VIKING survey \citep[][]{Fleuren12}. For SED fitting
analysis we use VIKING flux densities estimated from aperture
photometry with an aperture radius of 2$^{\prime\prime}$ for SDP.9
and SDP.11, 1$^{\prime\prime}$ for SDP.17, and 4$^{\prime\prime}$
for SDP.81 and SDP.130. Associated errors are derived from the
distribution of the flux
densities values that were obtained by taking  
aperture photometry at random positions in the field (avoiding
the region around detected sources).  \\

For SDP.81 and SDP.130, near-IR imaging data at 3.6 and 4.5$\,\mu$m
are also available from {\it Spitzer} \citep{Hop11}. At those
wavelengths the emission from the lens and the background galaxy are
comparable (i.e. source to lens flux density ratio $\gsim$ 0.2) and the separation of the two contributions was performed
by using the information from the SMA and the Keck data as a prior
(see Hopwood et al. for details). \\

Imaging data at 3.4, 4.6, 12 and 22\,$\mu$m, with an angular
resolution of 6.1$^{\prime\prime}$, 6.4$^{\prime\prime}$,
6.5$^{\prime\prime}$ and 12.0$^{\prime\prime}$ respectively, are provided by the
Wide-field Infrared Survey Explorer (WISE) \citep{WISE10} all sky survey. The WISE
images have a 5$\,\sigma$ photometric sensitivity of 0.068, 0.098, 0.86
and 5.4 mJy, respectively, in un-confused regions. 
Postage stamp images centred at the position of the five H-ATLAS/SDP lenses are shown in
Fig.\,\ref{fig:vik_wise}. All our targets are detected by the WISE at
3.4\,$\mu$m (W1) and 4.6\,$\mu$m (W2) while at 12\,$\mu$m (W3) and
22\,$\mu$m (W4) only SDP.9, SDP.17 and SDP.17 have a counterpart in the
WISE catalogue. In the following we adopt the WISE flux densities
determined by standard profile fitting\footnote{The {\it w$?$mpro} photometry in the Wise All Sky Data Release catalogue, with $?$ equal
to 1, 2, 3 or 4 depending on the observing band.} as all our
targets have extended source flag {\it text$\underline{~}$flg}\,=\,0. For SDP.81 and
SDP.130 we use the available 95\% upper limit at 12 and 22\,$\mu$m.

\subsection{Spectroscopic redshifts}

For all our targets the redshift of the background galaxy has been
constrained through the detection of CO emission lines by Z-spec
\citep{Lupu12}, GBT/Zpec \citep{Fra11,Harris12}, PdBI
\citep[N10,][]{Om11,Om13, George14} and CARMA (Leew et al. in
prep.). H$_{2}$O was detected in SDP.17 \citep{Om11}, SDP.9 and SDP.81
\citep{Om13} with PdBI, while emission from [CII] and [OIII] has been measured in SDP.81 \citep{Val11}. 
Optical spectra of the foreground galaxy were taken with the William
Herschel Telescope (WHT) for SDP.11 and SDP.17 and with the Apache
Point Observatory 3.5-meter telescope for SDP.130 (N10), giving
spectroscopic redshifts in the range $z_{\rm spec}=0.22-0.94$. 
For SDP.81
an optical spectrum was already available via SDSS, which gives
$z_{\rm spec}=0.299$. SDP.9 has an optical spectroscopic redshift
$z_{\rm spec}=0.613$ recently obtained with the Gemini-South telescope
\citep{Buss13}. A summary of available photometric and spectroscopic
information is given in Table\,\ref{tab:phot_param}.

\begin{table*}
\vspace{-0.3cm}
 \caption{Photometric data, spectroscopic redshifts and best-fit
   SED parameters for the lens and for the background source. At those
   wavelengths where the separation between the foreground galaxy and the
   background source was not possible, the total (lens+source)
   photometry is provided. All the errors 
   correspond to the 68 per cent confidence interval. Unless otherwise indicated, the data come from N10.}
\vspace{-0.2cm}
 \label{tab:phot_param}
 \centering
 \scriptsize
 \begin{tabular}{@{}lccccc}
  \hline
  \hline
    &   SDP.9  &   SDP.11   &   SDP.17  &  SDP.81  &  SDP.130  \\
  \hline
IAU name  & 
J090740.0$-$004200   &
J091043.1$-$000321   &
J090302.9$-$014127   &
J090311.6+003906       &
J091305.0$-$005343   \\
  \hline
\multicolumn{6}{l}{\bf LENS} \\
%
 Keck $g$ ($\mu$Jy)    &
 1.50$\pm$0.23   &
 1.54$\pm$0.20      &
 ...      &
  66.0$\pm$14    &
 18.4$\pm$2.7     \\
 Keck $i$ ($\mu$Jy)    &
 21.5$\pm$2.6   &
 23.8$\pm$1.9       &
 ...      &
 105$\pm$21    &
 93.7$\pm$0.9     \\
 SDSS\,$u$ ($\mu$Jy)    &
 0.24$\pm$0.23    &
 0.57$\pm$0.58        &
 ...      &
 3.9$\pm$2.0      &
 1.7$\pm$1.7      \\
 SDSS\,$g$ ($\mu$Jy)    &
 1.79$\pm$0.43    &
 1.01$\pm$0.45        &
 ...      &
 24.9$\pm$1.1      &
 19.4$\pm$0.7      \\
 SDSS\,$r$ ($\mu$Jy)    &
 5.81$\pm$0.70   &
 3.94$\pm$0.65       &
 ...      &
 115$\pm$2      &
 66.1$\pm$1.2      \\
 SDSS\,$i$ ($\mu$Jy)    &
 14.9$\pm$1.1      &
 11.3$\pm$1.0      &
 ...      &
 198$\pm$4      &
 109$\pm$2      \\
 SDSS\,$z$ ($\mu$Jy)    &
 27.0$\pm$3.7      &
 ...      &
 ...      &
 278$\pm$8      & 
 143$\pm$7      \\
 HST/F110W  ($\mu$Jy)     &
 37.4$\pm$1.6                &
 34.6$\pm$1.5      &
 13.2$\pm$1.0      &
 273$\pm$4      &
 202$\pm$61      \\
 HST/F160W  ($\mu$Jy)     &
 60.3$\pm$3.0                &
 54.4$\pm$2.9      &
 19.8$\pm$2.0      &
 381$\pm$8      &
 275$\pm$83     \\
 VIKING\,$Z$  ($\mu$Jy)          &
 31.3$\pm$1.6      &
 ...      &
 ...      &
 210$\pm$2      &
 157$\pm$2      \\
 VIKING\,$Y$  ($\mu$Jy)          &
 33.0$\pm$4.3     &
 ...      &
 ...      &
 233$\pm$5      &
 196$\pm$3     \\
 VIKING\,$J$  ($\mu$Jy)          &
 52.0$\pm$4.0                    &
 ...      &
 ...      &
 379$\pm$5      &
 244$\pm$5     \\
 VIKING\,$H$  ($\mu$Jy)          &
 ...      &
 ...      &
 ...      &
 485$\pm$8      &
 310$\pm$9      \\
 VIKING\,$Ks$  ($\mu$Jy)          &
 ...      &
 ...      &
 ...      &
 630$\pm$12      &
 388$\pm$9      \\
 {\it Spitzer}\,3.6$\mu$m  ($\mu$Jy)$^{(b)}$          &
 ...      &
 ...      &
 ...      &
 354$\pm$43      &
 213$\pm$30      \\
 {\it Spitzer}\,4.5$\mu$m  ($\mu$Jy)$^{(b)}$          &
 ...      &
 ...      &
 ...      &
 220$\pm$40      &
 230$\pm$10      \\
%
 Redshift   &  
 0.6129$\pm$0.0005$^{(a)}$  & 
 0.7932$\pm$0.0012             &
 0.9435$\pm$0.0009             &
 0.2999$\pm$0.0002             &
 0.2201$\pm$0.002       \\
 $M_{\star}$ ($10^{10}$\,$M_{\odot}$)   &
 6.8$_{-1.6}^{+1.4}$          &  
 10.1$_{-2.5}^{+2.8}$          &  
 3.9$_{-1.3}^{+1.6}$          &
 10.3$_{-2.8}^{+2.8}$          &
 4.2$_{-1.1}^{+1.0}$          \\
 SFR ($M_{\odot}$\,yr$^{-1}$)              & 
 0.19$_{-0.10}^{+0.13}$      &     
 0.77$_{-0.37}^{+0.46}$        &
 3.3$_{-1.7}^{+1.9}$        &
 0.25$_{-0.16}^{+0.28}$    &
 0.06$_{-0.05}^{+0.09}$    \\
S\'{e}rsic index at F110W ($n_{\rm S}^{\rm F110W}$)    & 
5.1             &
1.0+2.8            &
0.7+11.0             &
 2.3+2.0            &
 multiple profiles      \\
S\'{e}rsic index at F160W ($n_{\rm S}^{\rm F160W}$)    & 
5.8            &
 1.0+4.5            &
 0.6+9.7            &
 2.9+0.9            &
 multiple profiles      \\
  \hline
\multicolumn{6}{l}{\bf BACKGROUND SOURCE} \\
%
 Keck\,$g$ ($\mu$Jy; 5$\sigma$ upper limits)  &
 $<$0.20    &
 $<$0.32    &
 ...      &
 $<$0.28      &
 $<$0.16$^{\dagger}$      \\
 Keck\,$i$ ($\mu$Jy; 5$\sigma$ upper limits)  &
 $<$0.75    &
 $<$1.2      &
 ...      &
 $<$0.87      &
 $<$0.53$^{\dagger}$      \\
 HST/F110W ($\mu$Jy)  &
 3.6$\pm$0.5      & 
 23.8$\pm$4.3    &
 8.7$\pm$1.7      &
 1.9$\pm$0.4      &
 11.1$\pm$3.3      \\
 HST/F160W ($\mu$Jy)  &
 12.3$\pm$1.4     &
  47.7$\pm$6.9    &
 16.2$\pm$3.2      &
 4.5$\pm$0.8      &
 27.1$\pm$8.1      \\
 {\it Spitzer}\,3.6$\mu$m  ($\mu$Jy)$^{(b)}$          &
 ...      &
 ...      &
 ...      &
 62$\pm$44      &
 44$\pm$20      \\
 {\it Spitzer}\,4.5$\mu$m  ($\mu$Jy)$^{(b)}$          &
 ...      &
 ...      &
 ...      &
 126$\pm$54      &
 47$\pm$10      \\
 WISE 12\,$\mu$m (mJy)  &
 1.31$\pm$0.05    &
 2.14$\pm$0.06    &
 1.37$\pm$0.05      &
 $<$0.26      &
 $<$0.49      \\
 WISE 22\,$\mu$m (mJy)  &
 4.2$\pm$0.3    &
 9.2$\pm$0.4    &
 5.6$\pm$0.4      &
 $<$1.76      &
 $<$2.68      \\
 PACS 70\,$\mu$m (mJy)  &
 ...    &
 ...      &
 ...      &
 $<$8.0$^{(f)}$      &
 $<$9.0$^{(f)}$      \\
 PACS 100\,$\mu$m (mJy)  &
 187$\pm$57    &
 198$\pm$55      &
 78$\pm$55      &
 $<$62      &
 ...      \\
 PACS 160\,$\mu$m (mJy)  &
 416$\pm$94    &
 397$\pm$90    &
 182$\pm$56      &
 51$\pm$5$^{(f)}$      &
 45$\pm$8$^{(f)}$      \\
 SPIRE 250\,$\mu$m (mJy)  &
 485$\pm$73    &
 442$\pm$67    &
 328$\pm$50   &
 129$\pm$20      &
 105$\pm$17     \\
 SPIRE 350\,$\mu$m (mJy)  &
 323$\pm$49    &
 363$\pm$55    &
 308$\pm$47      &
 182$\pm$28      &
 128$\pm$20      \\
 SPIRE 500\,$\mu$m (mJy)  &
 175$\pm$28    &
 238$\pm$37    &
 220$\pm$34      &
 166$\pm$27      &
 108$\pm$18      \\
 SMA 880\,$\mu$m (mJy)$^{(a)}$  &
 24.8$\pm$3.3    &
 30.6$\pm$2.4    &
 54.7$\pm$3.1      &
 78.4$\pm$8.2      &
 36.7$\pm$3.9      \\
 MAMBO 1200\,$\mu$m (mJy)  &
 7.6$\pm$1.4        &
 12.2$\pm$2.3      &
 15.3$\pm$3.9      &
 20.0$\pm$3.1      &
 11.2$\pm$2.1      \\
%
%
 Redshift   &
 1.577$\pm$0.008               &
 1.786$\pm$0.005               &
 2.3049$\pm$0.0006$^{(c)}$  &
 3.042$\pm$0.001               &
 2.6260$\pm$0.0003          \\
$\mu^{(d)}$  &
6.29$_{-0.26}^{+0.27}$    &
7.89$_{-0.25}^{+0.21}$    &
3.56$_{-0.17}^{+0.19}$    &
10.6$_{-0.7}^{+0.6}$       &
3.09$_{-0.17}^{+0.19}$    \\
$\mu_{\rm SMA}^{(a)}$  &
8.8$\pm$2.2      &
10.9$\pm$1.3    &
4.9$\pm$0.7      &
11.1$\pm$1.1    &
2.1$\pm$0.3      \\
 $L_{\rm dust}/\mu$ (10$^{12}$\,$L_{\odot}$)  &
 7.4$_{-1.2}^{+1.2}$      &
 7.6$_{-1.0}^{+1.2}$      &
 20.7$_{-3.5}^{+4.0}$    &
 5.1$_{-0.7}^{+0.8}$      &
 9.2$_{-1.4}^{+1.6}$      \\
 SFR$/\mu$ ($M_{\odot}$\,yr$^{-1}$)  &
 366$_{-259}^{+441}$    &
 650$_{-456}^{+157}$      &
 2325$_{-485}^{+472}$     &
 527$_{-91}^{+102}$         &
 1026$_{-206}^{+318}$      \\
 $T_{\rm dust}^{\rm (warm)}$ (K)  &
 45.4$_{-2.2}^{+2.8}$     &
 46.9$_{-2.5}^{+2.8}$     &
 43.3$_{-3.4}^{+5.9}$     &
 39.3$_{-1.5}^{+2.1}$      &
 37.3$_{-1.6}^{+2.1}$     \\
%
%
 $\xi_{\rm dust}^{\rm (warm)}$  &
 0.80$_{-0.10}^{+0.10}$     &
 0.84$_{-0.08}^{+0.07}$    &
  0.72$_{-0.07}^{+0.06}$   &
 0.74$_{-0.10}^{+0.10}$    &
 0.73$_{-0.11}^{+0.11}$    \\
 $M_{\rm dust}/\mu$ (10$^{8}$\,$M_{\odot}$)  &
 6.7$_{-1.0}^{+1.2}$     &
 6.6$_{-0.9}^{+1.1}$     &
 28.8$_{-4.9}^{+7.0}$    &
 10.6$_{-1.9}^{+2.9}$      &
 22.7$_{-4.1}^{+5.8}$     \\
 $M_{\star}/\mu$ (10$^{10}$\,$M_{\odot}$)  &
 7.1$_{-2.3}^{+4.2}$      &
 18.7$_{-4.5}^{+5.8}$    &
 24.2$_{-4.0}^{+8.6}$   &
 6.6$_{-1.9}^{+2.6}$      &
 13.7$_{-2.5}^{+3.8}$      \\
 $M_{\rm gas}/\mu$ (10$^{10}$\,$M_{\odot}$)$^{(e)}$  &
 3.4$_{-0.9}^{+1.2}$       &
 3.0$_{-0.8}^{+1.1}$       &
 5.9$_{-1.6}^{+2.2}$       &
 3.3$_{-0.9}^{+1.2}$       &
 5.3$_{-1.4}^{+2.0}$      \\
 $f_{\rm gas}=M_{\rm gas}$/($M_{\rm gas} + M_{\star}$) &
 0.32$_{-0.10}^{+0.13}$       &
 0.14$_{-0.04}^{+0.06}$       &
 0.19$_{-0.05}^{+0.06}$       &
 0.33$_{-0.09}^{+0.09}$       &
 0.28$_{-0.07}^{+0.08}$       \\
%
%
 $\tau_{\rm gas}=M_{\rm gas}$/SFR (Myr) &
 103$_{-64}^{+236}$      &
 57$_{-23}^{+99}$     &
 26$_{-8}^{+11}$       &
 63$_{-19}^{+27}$      &
 50$_{-18}^{+23}$     \\
 \hline
\multicolumn{6}{l}{\bf LENS + BACKGROUND SOURCE} \\
%
 Keck $g$ ($\mu$Jy)    &
 ...   &
 ...       &
 1.15$\pm$0.23      &
 ...     &
 ...     \\
 Keck $i$ ($\mu$Jy)    &
 ...   &
 ...       &
 9.31$\pm$1.9      &
 ...     &
 ...     \\
 SDSS\,$u$ ($\mu$Jy)    &
 ...      &
 ...      &
 3.3$\pm$1.6      &
 ...      &
 ...      \\
 SDSS\,$g$ ($\mu$Jy)    &
 ...      &
 ...      &
 3.9$\pm$0.6      &
 ...      &
 ...      \\
 SDSS\,$r$ ($\mu$Jy)    &
 ...      &
 ...      &
 7.7$\pm$1.0      &
 ...      &
 ...      \\
 SDSS\,$i$ ($\mu$Jy)    &
 ...      &
 ...      &
 15.3$\pm$1.5      &
 ...      &
 ...      \\
 SDSS\,$z$ ($\mu$Jy)    &
 ...      &
 21.5$\pm$4.2       & 
 11.8$\pm$6.0      &
 ...    & 
 ...      \\
 VIKING\,$Z$  ($\mu$Jy)          &
 ...      &
 40.1$\pm$0.5      &
 10.5$\pm$0.5      &
 ...      &
 ...      \\
 VIKING\,$Y$  ($\mu$Jy)          &
 ...      &
 53.5$\pm$1.1      &
 17.1$\pm$1.1      &
 ...      &
 ...      \\
 VIKING\,$J$  ($\mu$Jy)          &
 ...      &
 78.4$\pm$1.1      &
 25.0$\pm$3.7      &
 ...      &
 ...      \\
 VIKING\,$H$  ($\mu$Jy)          &
 92.4$\pm$5.2        &
 120.0$\pm$2.8      &
 34.9$\pm$3.9       &
 ...      &
 ...      \\
 VIKING\,$Ks$  ($\mu$Jy)          &
 123.2$\pm$5.4      &
 199.7$\pm$2.4      &
 74.5$\pm$4.8       &
 ...      &
 ...      \\
 WISE 3.4\,$\mu$m ($\mu$Jy)  &
 218.2$\pm$3.9    &
 519$\pm$6.9      &
 132.3$\pm$3.9      &
 343.2$\pm$5.1     &
 208.0$\pm$4.3      \\
 WISE 4.6\,$\mu$m ($\mu$Jy)  &
 300.6$\pm$6.6    &
 632$\pm$10      &
 209.1$\pm$6.8      &
 303.9$\pm$6.9      &
 233.8$\pm$7.7      \\
  \hline
  \hline
\multicolumn{6}{l}{\scriptsize $^{(a)}$ from \cite{Buss13}; $^{(b)}$
  from \cite{Hop11}; $^{(c)}$ from \cite{Om13}; $^{(d)}$ from \cite{Dye13}.} \\
\multicolumn{6}{l}{\scriptsize $^{(e)}$ from \citet[][Table\,1, assuming a 30\% error]{Fra11} and
  \citet[][Table\,4, assuming a 30\% error]{Lupu12}; $^{\dagger}$
  ``tentative''} \\
 \end{tabular}
\footnotetext{ciccio}
\end{table*}


\section{Lens subtraction}

According to Fig.\,\ref{fig:HST_images1} and
Fig.\,\ref{fig:HST_images2} (left panels), the HST data alone strongly support the idea of a gravitational lensing event in three of the five targets, namely SDP.9, SDP.11 and SDP.17, through the detection of a diffuse ring-like structure around a central elliptical galaxy.
Hints of lensing are also found in the WFC3 images of SDP.81, where a faint arclet is visible $\sim$1.5 arcseconds away from the central elliptical galaxy in the west direction. 
For SDP.130 no clear evidence of gravitational lensing can be claimed from the HST
images alone, where the system resembles a lenticular galaxy. \\

In order to unveil the full morphology of the lensed source, the light profile of the foreground galaxy needs to be fitted and subtracted. We use the GALFIT software \citep{Peng02} to construct models of the light profiles for each lensing system. GALFIT performs a non-linear 2D minimisation and allows multiple profiles to be simultaneously fitted.
As these lensing systems are photometrically blended in the HST data,
in order to achieve a good fit to the lens galaxy it is necessary to
fit profiles to both the lens and source components in the same
model. Once a satisfactory model is achieved for the whole system, only the
best fit lens profile is then subtracted. If there are other
sources within the fitting region they are either masked or, if close
enough to the main source to cause significant photometric blending,
are included in the fit (e.g. the edge-on galaxy at the
north-west side of SDP.11; Figs\,\ref{fig:HST_images1}-\ref{fig:HST_images2}). Where available,
sub-arcsecond resolution ancillary data (e.g. from the SMA) are used
to guide the fitting process. For each image, nearby stars were
combined to give an empirical Point Spread Function (PSF). All star
candidates were checked for saturation, normalised and re-centred
before being median combined. For SDP.11 only one suitable star is
available. \\

For each image initially one S\'{e}rsic profile was fitted to the
foreground lens in order to gauge the level of lensed structure above
the detection limit. Then each GALFIT model is built up by adding
extra profiles until both the lens and source galaxy components are
well represented. The process is iterative and follows the basic loop
of applying GALFIT, inspecting the results, adjusting the parameters
and possibly adding more complexity where necessary before re-applying
GALFIT. This is a process that relies on thorougher visual inspection
at each stage, with comparison to other available data, such as the
SMA data to check the profiles/model associated with the lensed
structure. The fitting process generally started with the higher
signal-to-noise ratio F160W image, and then these results
used as a prior for the initial guess for F110W. A close eye was
kept to try and maintain a reasonable similarity in the profile
orientation and ellipticity for both bands, where that was possible.
The resulting lens-subtracted images are shown in the right panels of
Fig.\,\ref{fig:HST_images2} and compared with the signal-to-noise
ratio contours at 880$\,\mu$m from the SMA \citep[N10;][]{Buss13}.  
Below we discuss the GALFIT results for the five sources individually.\\
\vskip-0.6cm
$~$ \\
{\bf SDP.9}. The foreground galaxy is fitted with a single S\'{e}rsic
profile of index $n_{\rm s}=\,$5.1 in F110W and $n_{\rm
  s}=\,$5.8 in F160W. The light profile is therefore consistent
with that of an elliptical galaxy. After the subtraction of the lens,
a diffuse ring-like structure is clearly revealed, particularly at
1.6$\,\mu$m. The ring contains two main knots of near-IR emission to
the north and south of the lens position and two fainter ones to the
east and west. \\
\vskip-0.6cm
$~$ \\
{\bf SDP.11}. This is the 500$\,\mu$m brightest lens candidate
selected in the H-ATLAS/SDP field (see Table\,\ref{tab:phot_param})
and even without the subtraction of the foreground galaxy it is clear
that the background source is lensed into an Einstein ring. The ring
is particularly elongated with a significant amount of
substructure, which suggests the presence of several clumps of
rest-frame UV/optical emission in the source plane (D13),
consistent with what was found for the H-ATLAS lensed galaxy
presented in \cite{Fu12}. The foreground galaxy required two S\'{e}rsic profiles in each of the
bands, where one profile is approximately an exponential disk ($n_{\rm
  s}\sim1$) and the other profile has index $n_{\rm s}=$2.8 at
1.1$\,\mu$m and $n_{\rm s}=$4.5 and 1.6$\,\mu$m. Also in this case the light profile is indicative of an elliptical/lenticular galaxy. \\
\begin{figure*}
\hspace{+1.0cm}
\makebox[\textwidth][c]{
\includegraphics[width=0.94\textwidth]{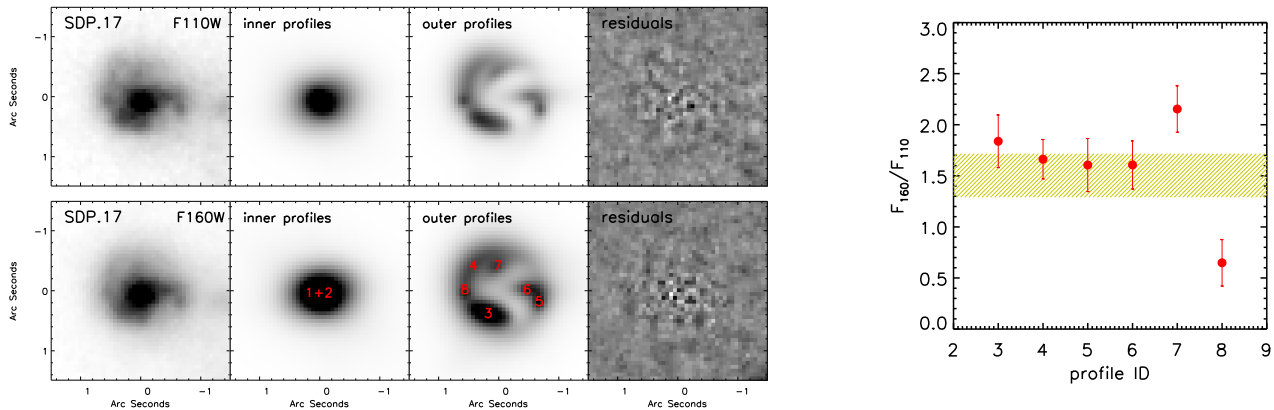}
} 
\vspace{-3.6cm}
\caption{GALFIT results for SDP.17 at 1.1$\,\mu$m (top panels) and at
 1.6$\,\mu$m (bottom panels). From left to right it is shown the
 input image, the model inner profiles (that we assume describe the
 lens), the model outer profiles, and the residuals.The 1.6- to
 1.1-$\mu$m flux density ratios of the outer profiles (marked by
 numbers in figure) are shown in the right panel and compared with the
 1.6- to 1.1-$\mu$m flux density ratio of the two inner
 profiles (yellow shaded region). If the outer profiles are part of
the lensed source then their near-IR  flux density ratio would increase
towards the edges of the image, thus reflecting the reddening of the
SED of the background galaxy, due to both high redshift and dust
extinction. This is only the case for profile 8. Therefore the
background source is assumed to be comprised of profiles $3-4-5-6-7$.}
\label{fig:SDP17_profiles}
\end{figure*}
\vskip-0.6cm
$~$ \\
{\bf SDP.17}. At first glance, this system resembles a face-on spiral galaxy with two prominent spiral arms.
However we know from spectroscopic follow-up observations that this
system has an optical redshift of 0.9435 (N10) and a redshift of 2.305
from the detection of both CO \citep{Lupu12,Harris12} and $H_{2}0$
\citep{Om11} lines, thus indicating the presence of two objects along
the same line of sight. Follow-up observations with the SMA
\citep{Buss13} show that the sub-mm/mm emission is relatively compact,
concentrated within $\sim0.6^{\prime\prime}$ from the centre of the
source, but fails to resolve the individual lensed images. \\
A satisfactory fit to the observed light distribution of this object
requires 8 profiles as illustrated in Fig.\,\ref{fig:SDP17_profiles}:
two accounts for the innermost region (i.e. that within
$\lsim0.3\,$arcsec from the centre), one with S\'{e}rsic index $n_{\rm
  s,1}\gsim10$ and the another one (less extended) with $n_{\rm
  s,2}\sim0.6$ (at both 1.1 and  1.6$\,\mu$m). We assume that these
two profiles describe the foreground galaxy (or at least most of it),
which is acting as a lens. The other 6 profiles may either be all
associated with the lensed source or, at least some of them, may belong to the
foreground galaxy. In order to understand the more likely scenario, we
have derived the  1.6-$\mu$m to 1.1-$\mu$m flux density ratio, $F_{\rm
  1.6}/F_{\rm 1.1}$, for each of the outermost profiles. In fact, if
the lens had spiral arms then we would expect the arms to display
bluer colours than the bulge and the ratio $F_{\rm 1.6}/F_{\rm 1.1}$
would decrease from the centre toward the outer regions of the
galaxy. On the contrary, if the spiral-arms-like structure  is part of
the lensed source then the same flux density ratio would increase
towards the edges of the image, thus reflecting the reddening of the
SED of the background galaxy, due to both high redshift and dust
extinction \citep[although examples of sub-mm selected galaxies comprising some
relatively ``blue'' components exist; see e.g.][]{Ivison10}. The measured flux density ratios are shown in
Fig.\,\ref{fig:SDP17_profiles} (right panel).  We find that the
profile labelled as 8 is significantly bluer than the lens. 
It might be either another foreground object, not necessarily associated with
the lens, or a small star forming region in the lens itself, which
could explain the detection of the lens in CO in the Z-spec spectrum
(N10). We exclude that it corresponds to a dust-free region in the source
plane; in fact, if that was the case, its lensed counter-image would have a similar
1.6-$\mu$m to 1.1-$\mu$m flux density ratio, but this is not the case.
Indeed all the other outer profiles have either redder colours than the lens
(e.g. profile 3 and profile 7) or colours similar to
it. Therefore, in this work and in D13 we assume that the lensed object is made up of profiles
$3-4-5-6-7$. The complicated clumpy structure of SDP.17 can be accounted for by a
lensing event characterised by two distinct knots of rest-frame UV/optical
emission in the source plane (D13). \\
\vskip-0.6cm
$~$ \\
{\bf SDP.81}. The fit to the Keck image performed in N10 required two
profiles: a compact elliptical S\'{e}rsic core plus a subdominant
exponential disk. Two S\'{e}rsic profiles are required to achieve a good lens subtraction in both HST bands, with indexes $n_{\rm
  s,1}=\,$2.3 and $n_{\rm s,2}=\,$2.0 for F110W, and $n_{\rm
  s,1}=\,$2.9 and $n_{\rm s,2}=\,$0.9 for F160W.
Once the lens is subtracted (Fig.\,\ref{fig:HST_images2}), two arclets, on opposite sides with respect to the centre of the lens, are revealed. The smaller one, on the west, was barely visible before the subtraction of the lens. The remarkable similarity of the residuals to the structure revealed in the sub-mm by the SMA (see fig.\,\ref{fig:HST_images2}) supports the robustness of the lens fitting procedure. \\
\vskip-0.6cm
$~$ \\
{\bf SDP.130}. This system was also fitted in N10 using two profiles: a compact elliptical S\'{e}rsic core plus an exponential disk.
In the HST images the exponential disc is now clearly resolved into
two diffuse spiral arms. The lens galaxy can thus be classified as an
Sa galaxy. The arm extending in the south-east direction reveals a
substructure oriented in an almost orthogonal direction to the arm
itself. This small structure produces some emission in the sub-mm (detected
with the SMA, Fig.\,\ref{fig:HST_images2}) and may suggest
that an interaction of the arm with another object is on-going and is
triggering some star-formation activity. Another structure is visible
close to the bulge, in the north-west direction, but is undetected by
the SMA. Its nature is unclear. Overall, the morphology of this system
in the HST images is not suggestive of a lensing event. This is
because, in the near-IR, the lensed object is masked by the prominent
bulge of the foreground galaxy. The subtraction of the lens is
therefore necessary to reveal the background source.  \\
SDP.130 presents the greatest challenge of the five lenses in terms of
subtracting off the lensing galaxy light profile. With the superior
resolution of the HST images it is now clear that spiral arms are
present, although possibly these have suffered some disruption via the
interaction with a smaller object (the one to the south east mentioned
before) and now form more of an elongated ring. In addition there may
be a bar structure across the bulge. The GALFIT model of the lens is
thus made up of multiple profiles fitted to the bulge, the bar, the spiral
arms and the small interacting object. The final model has sixteen
fitted profiles in all, five associated to the lensing galaxy core,
where three profiles represent the bulge and two the bar `ends'. As
shown in Fig.\,\ref{fig:HST_images2}, the
two profiles representing the lensed structure correspond well to the
two most prominent knots seen in the SMA data. Overall SDP.130 is well
subtracted. \\

\section{Spectral energy distributions}

%
%
%
\begin{figure*}
\begin{centering}
\vspace{+0.5cm}
\hspace{+4.5cm}
\makebox[\textwidth][c]{
\includegraphics[width=1.4\textwidth]{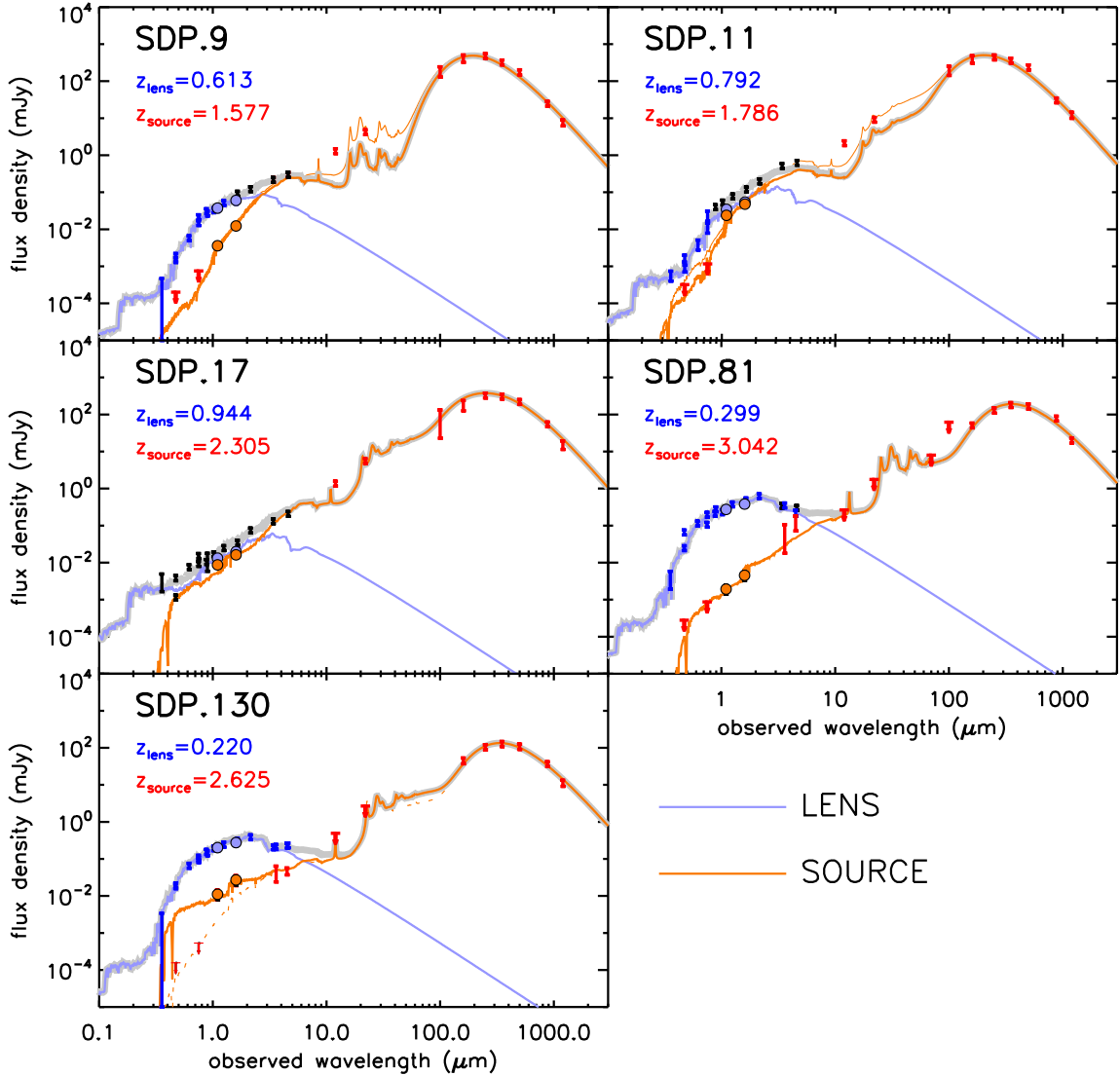}
} 
\vspace{+0.0cm}
\caption{Spectral Energy Distributions of the lens and of the
  background source for the five H-ATLAS/SDP gravitational lensing
  systems. The new photometric data points from HST/WFC3 are indicated
  by dots (cyan for the lens and orange for the background source)
  while other existing photometric data are represented with either
  error bars or downward arrows in case of upper limits. The optical
  data are from SDSS and Keck while measurements at near/mid-IR are
  from VIKING, WISE, and {\it Spitzer}. The sub-mm/millimetre
  photometry is from PACS/{\it Herschel}, SPIRE/{\it Herschel}, SMA
  and MAMBO/IRAM. Upper limits at PACS/Herschel wavelengths are shown
  at 3$\,\sigma$. Data points are blue for the lens photometry, red
  for the background source photometry and black for the
  lens+source photometry. The best-fit SED is in cyan for the lens and
  in orange for the source. The thick grey line is their sum. For
  SDP.9 and SDP.11 the lighter orange curve show the best-fit results
  for the lensed source when the WISE data points at 12 and 22$\,\mu$m
  are included in the fit. For SDP.130, the dashed curve is the
  best-fit SED obtained for the lensed source  when the Keck upper
  limits are also taken into account.}
\label{fig:SEDs}
\end{centering}
\end{figure*}

The GALFIT decomposition allows us to measure the photometry of the
lensing galaxy and of the background source separately. We use
aperture photometry on the GALFIT best-fit model images to derive the
flux densities at 1.1 and 1.6$\,\mu$m. Photometric errors are obtained
by taking aperture photometry of the sky (with the same aperture
radius used to measure the flux of the targets) at random positions
and estimating the corresponding rms. The results are listed in Table\,\ref{tab:phot_param} and shown in Fig.\,\ref{fig:SEDs} (coloured dots: orange for the background source and cyan for the foreground galaxy) together with other available photometric data. \\

We fit the observed SEDs using the public code
Multiwavelength Analysis of Galaxy Physical Properties
\cite[{\sc magphys};][]{daCunha08}, which exploits a large library of optical and IR templates linked together in a physically consistent way. 
The evolution of the dust-free stellar emission is computed using the
population synthesis model of \cite{BC03}, by assuming a \cite{Chab03} initial mass function (IMF) that is
cutoff below 0.1 and above 100 M$_{\odot}$; adopting a Salpeter IMF
instead gives stellar masses that are a factor of $\sim$1.5 larger. \\
The attenuation of starlight by dust is described by the two-component
model of \cite{CF00}, where dust is associated with the birth clouds and with the diffuse interstellar medium (ISM).
Starlight is assumed to be the only significant heating source
(i.e. any contribution from an active galactic nucleus is neglected).
The dust emission at far-infrared to sub-mm/millimetre wavelengths
is modelled as a two modified grey-body SED with dust emissivity index
$\beta=1.5$ for the warm dust (30$-$60\,K) and $\beta=2$ for the cold
dust (15$-$30\,K). The dust mass absorption coefficient, $k_{\lambda}\propto\lambda^{-\beta}$, is approximated
as a power law with normalisation $k_{\rm 850\mu m}=0.077$\,m$^{2}$\,kg$^{-1}$ \citep{D00}. \\
Among the best SED-fit parameters provided by {\sc magphys} we report the
following in
Tab.\,\ref{tab:phot_param}: total ($3-1000\,\mu$m) IR luminosity of dust
emission ($L_{\rm dust}$ or $L_{\rm IR}$),  star formation rate
(averaged over the last 100\,Myr), SFR, stellar mass, $M_{\star}$,
dust mass, $M_{\rm dust}$, temperature of the warm dust component,
$T_{\rm dust}^{\rm (warm)}$, fraction of the IR luminosity due to the warm dust,
$\xi_{\rm dust}^{\rm (warm)}=L_{\rm dust}^{\rm (warm)}/L_{\rm
  dust}$. 
In order to derive the {\it intrinsic} properties of the background source,
a correction for the amplification due to lensing is
applied. Thousands of simulated
values for the {\it observed} parameters are generated from the likelihood
probability distributions provided by {\sc magphys} and then divided
by the magnification factors randomly drawn
from a Gaussian distribution with mean
value and rms taken from D13 (and also reported in
Table\,\ref{tab:phot_param}).  The medians of the the simulated
amplification-corrected values are taken as the best
estimates of the {\it intrinsic} properties of the source and are those listed in
Tab.\,\ref{tab:phot_param}. The associated
errors correspond to the confidence interval in the 16th to 84th percentile
range.  \\

For the fit to the SED of the background source we adopt the SED
templates calibrated to reproduce the ultraviolet-infrared SEDs of
local, purely star-forming ULIRGs \citep{daCunha10}, while we use
dust-free SED templates to fit the SED of the lenses (i.e. pure
Bruzual \& Charlot 2003 models). For the latter we just report the
estimated mass in stars and star formation rate in
Tab.\,\ref{tab:phot_param}. \\
In general, we assume that the measured SDSS and VIKING photometry have
contributions from both the foreground galaxy and the lensed source,
unless otherwise stated. In fact, ground based observations are
limited by the seeing, which makes it extremely difficult to separate
the lens from the background source in our relatively compact
targets. We further assume that the emission at 12 and 22$\,\mu$m (as
measured by WISE) is entirely contributed by the lensed source  while
the WISE photometry at 3.4 and 4.6$\,\mu$m carries contributions from
both the lens and the source, unless otherwise stated. The best-fit SED models are shown Fig.\,\ref{fig:SEDs} while the corresponding best-fit parameters are listed in Table\,\ref{tab:phot_param}.  Below we provide more details on the fit to the SED for each object individually. \\
\vskip-0.6cm
$~$ \\
{\bf SDP.9}. At wavelengths $\lambda\lsim1\,\mu$m the emission is
dominated by the foreground galaxy. The flux density ratio between the source and
the lens increases from 0.08 to 0.2 going from 1.1\,$\mu$m to 1.6\,$\mu$m,
so that we expect the $H$ and $Ks$ VIKING photometry to carry significant contributions from both the
lens and the background source. Therefore, for the SED of
the foreground galaxy we have adopted the SDSS and the $Z+Y+J$ VIKING
photometry, as well as the HST lens photometry. For the background
source, we have fitted the corresponding HST photometry, together with
$5\,\sigma$ upper limits from Keck in the optical (N10) and all the
available data at mid-IR (i.e. WISE W3+W4) to sub-mm/mm wavelengths,
where the contribution form the lens is null. All the other
photometric data are used as upper limits in the fit.
The results are shown in the top-left panel of Fig.\,\ref{fig:SEDs}
and are found to be independent on the inclusion of the Keck upper
limits. However we fail to reproduce the WISE data point at 12$\,\mu$m
(light orange curve). There is a clear excess at mid-IR wavelengths
that may be due to emission from a dusty torus around an Active
Galactic Nucleus (AGN). In fact, the presence of a dust-obscured AGN
in SDP.9 is suggested by the analysis of \cite{Om11} on the
H$_{2}$O(2$_{02}$−1$_{11}$)/CO(8-7)  and I(H$_{2}$O)/L$_{\rm FIR}$
ratios.  Our SED models do not include any AGN component, which may
provide the dominant contribution to the continuum mid-IR emission. Therefore we assume as our best-fit SED model the one derived by ignoring the WISE W3+W4 data points (thick orange curve). 
\\
\vskip-0.6cm
$~$ \\
{\bf SDP.11}. The foreground galaxy and the lensed source have about
the same flux densities at near-IR wavelengths. This means that  lower
spatial resolution near-IR photometric data, as those provided by the
VIKING survey, may carry similar contributions from the lens and the background source, although the two are completely blended together. 
Based on the available upper limits at optical wavelengths for the
lensed source we decided to use the $u+g+r+i$ SDSS photometry as well as
the HST lens photometry to describe the SED of the foreground galaxy,
while the fit to the SED of the lensed galaxy is performed on the
$Keck$ upper limits, the HST
source photometry and on photometric data at wavelengths
$>10\,\mu$m. However, also in this case we fail to reproduce the WISE
W3 data point (light orange curve). We conclude that a significant fraction of the mid-IR emission
in SDP.11 may come from an AGN. Also in this case we assume as our best-fit SED model the one derived by ignoring the WISE W3+W4 data points (thick orange curve). \\
\vskip-0.6cm
$~$ \\
{\bf SDP.17}. This case is similar to that of SDP.11 with the
foreground galaxy and the lensed source having very similar flux
densities at near-IR wavelengths. Therefore we fit the SED of the lens
including just the HST lens photometry and adopting the SDSS data
points as upper limits. We fit similarly for the lensed source with
the addition of the photometric data at wavelengths longwards of
10$\,\mu$m. No indication of a mid-IR ``excess'' is found in this case.\\
\vskip-0.6cm
$~$ \\
{\bf SDP.81}. The foreground galaxy completely dominates the emission
at wavelengths shorter than few $\mu$m. Therefore the fit to the SED
of the lens is done on the SDSS and VIKING data, as well as on the
lens photometry from HST and {\it Spitzer} \citep{Hop11}. As for the
background galaxy, we fit upper limits from Keck, source photometry
from HST and {\it Spitzer} and all the other available photometric
data above 10$\,\mu$m. \\ 
$~$ \\
{\bf SDP.130}. The lensed galaxy is an order of magnitude fainter
than the foreground galaxy at 1.1 and 1.6$\,\mu$m. On the other hand
the complicated morphology of the foreground galaxy may suggest that
the upper limits available at optical wavelengths for the lensed
source are poorly constrained. In fact, we failed to reproduce
simultaneously those limits and the HST photometry (dashed curve), as the increase in
flux density from 0.7 to 1.1$\,\mu$m is too steep. Therefore we also
show in fig.\,\ref{fig:SEDs} the best-fit SED model derived when the
Keck upper limits are not included in the fit (thick orange curve). The latter is assumed
as our best-fit SED model for the background galaxy.

\begin{figure}
\hspace{-4.5cm}
\makebox[\textwidth][c]{
\includegraphics[width=1.2\textwidth]{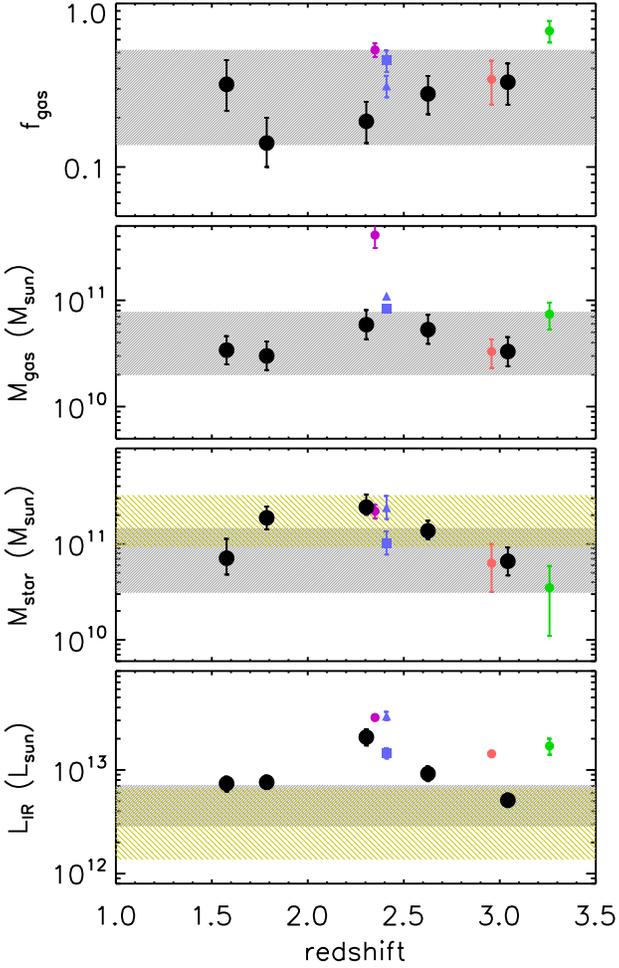}
} 
\vspace{-0.2cm}
\caption{Infrared luminosity, mass in stars, mass in molecular gas and
  gas fraction of the five H-ATLAS/SDP lensed galaxies (black dots) compared with
  other sub-mm selected lensed/un-lensed galaxies from literature: Ivison et
  al. (2013; cyan), Fu et al. (2012; green), Conley et al. (2011;
  red), Fu et al. (2013; purple). The shaded grey region corresponds to the 16th to 84th percentile
range of the distribution of values derived for the sample of
sub-millimeter galaxies with CO line measurements compiled by Bothwell
et al. (2013; with mass in stars taken from Hainline et al. 2011). The
shaded yellow region shows the infrared luminosities and masses in
stars estimated by Micha{\l}lowski et al. (2010; their values of $M_{\star}$ have
been rescaled by a factor of 1.8 to convert from Salpeter to 
Chabrier initial mass function.)}
 \label{fig:parameters}
\end{figure}

\section{Discussion}\label{discussion}

A detailed lens modelling of the lens-subtracted HST images, using a
multiwavelength semilinear inversion technique \citep{WD03}, is
presented in an accompanying paper (D13). The lens modelling provides
estimates of the magnification experienced by the background source
(at near-IR wavelengths) as well as constraints on both the total mass
distribution of the lens and the distribution of the UV/rest-frame
optical mission in the source plane. Here we use the D13 magnification
factors, also reported in Table\,\ref{tab:phot_param}, to derive the
{\it intrinsic} properties of the background sources as estimated from
the fit to their SED. In doing this we neglect the effect of
differential magnification, i.e the dependence of the magnification
factor on the observing wavelength \citep{Serjeant12}. In fact, the
amplification factors derived by D13 are consistent with those derived
for the same objects from the SMA images at 880$\,\mu$m \citep{Buss13}.
The values of the best-fit SED parameters are listed in
Table\,\ref{tab:phot_param} and shown in
Fig.\,\ref{fig:parameters}. \\

All the galaxies in the sample are classified as Ultra Luminous Infrared Galaxies
(ULIRGs, $10^{12}\,L_{\odot}\le L_{\rm IR}<10^{13}\,L_{\odot}$) with the exception of
SDP.17; its infrared luminosity, $L_{\rm IR}\sim 2\times10^{13}\,L_{\odot}$,
makes it an Hyper Luminous Infrared Galaxy (HyLIRGs). The latter
comprises two distinct objects in the
source plane, each one of $2-3$\,kpc in size, and separated by
a few kpc (Fig.1 of D13). This morphology may be indicative of an on-going
merger. Also the lens modeling of SDP.11 reveals multiple emitting regions in the source
plane, distributed over a region of $3-4$\,kpc, while for the other
systems one single object is required to reproduce the lensed
morphology in the HST images.  \\
SDP.17 is not
the first example of HyLIRG discovered among the H-ATLAS lens
candidates. Other examples are the $z=4.243$ lensed galaxy
analyzed by \cite{Cox11} and further investigated by \cite{Buss12},
the $z=3.259$ source lensed by a galaxy group discussed in
\cite{Fu12},  and the two starbursting galaxies (one of which is
weakly lensed) at $z=2.41$ presented
by \cite{Ivison13}.  Few more examples have been found in the {\it
  Herschel} Multi-tiered Extragalactic
Survey\citep[HerMES;][]{Oliver12}: a $z=2.9575$ source lensed by a galaxy group
\citep{Con11} and a weakly lensed merging system at $z=2.308$
\citep{Fu13}. \\

The inferred star formation
rates are in the range
$\gsim400-2000$\,M$_{\odot}$\,yr$^{-1}$, reaching a maximum for
SDP.17 and SDP.130. The derived dust masses, $M_{\rm dust}\sim(7-30)\times10^{8}\,$M$_{\odot}$ and dust temperatures,
$T=37-47\,$K, are in agreement with what is commonly found for high redshift ULIRGs/HyLIRGs
\citep{Michalowski10,Buss13}. At these high rates of star formation, the
mass in stars grows vey rapidly as the available molecular gas is quickly exhausted. With the aid of the new HST photometry we estimate that a
mass of $\sim(7-20)\times10^{10}\,M_{\odot}$ is already locked up in
stars. Although high, these values are a factor
$\times4$ lower than those derived by \cite{Hop11} for SDP.81 and
SDP.130. In fact, their SED fitting
could only rely on upper limits for the flux density of the lensed source at
wavelengths $\lambda<3.6\,\mu$m. 
Our estimates of the mass in stars are consistent with
those derived for other 
sub-millimeter selected  galaxies \citep[][see also
Fig.\,\ref{fig:parameters}]{Michalowski10,Hainline11,Yun12}. \\

For how long will these galaxies continue to
form stars? 
This depends on the mass of molecular gas,
$M_{\rm gas}$, still
available in these sources, which is information provided by  \cite{Fra11} and
\cite{Lupu12}  via the
detection of carbon monoxide (CO) emission
lines.
Here we have updated their estimates in light of the new
amplification factors derived by D13. The results are shown in
Table\,\ref{tab:phot_param} together with the derived
molecular gas
fraction, $f_{\rm gas}=M_{\rm gas}/(M_{\rm gas} + M_{\rm star})$ and
gas depletion time scale, $\tau_{\rm gas}=M_{\rm gas}/{\rm SFR}$. We find large reservoir of molecular
gas, $M_{\rm gas}>3\times10^{10}\,M_{\odot}$, consistently with
what is observed in other sub-millimeter galaxies \citep[][in Fig.\,\ref{fig:parameters}]{Bothwell13}. If star formation
is sustained at the rate estimated here, the gas will
be exhausted in less than 100\,Myr \citep[$\times2$ longer if gas
recycling is accounted for in stellar evolution;][]{Fu13}. By the end of this intense episode of
star formation such galaxies will have assembled a mass in stars of
$(1-3)\times10^{11}\,M_{\odot}$. Unless further gas is accreted from the surrounding environment
or through minor/major mergers they will passively evolve
 into massive ellipticals at the present time.
These galaxies are thus proto-ellipticals caught during their major episode
of star formation.

\section{Conclusions}

We have presented deep HST/WFC3 F110W+F160W follow-up observations
of the first gravitational lensing systems discovered by H-ATLAS in the
Science Demonstration Phase. The exquisit angular resolution of the
HST images has allowed us to resolve an Einstein ring in two of
these systems, and to identify multiple images in the others after a careful removal
of the foreground galaxy. The lens-subtrated images have been used to
model the rest-frame UV/optical emission in the source plane (D13) and
to improve the constraints on the mass in stars and on other physical
properties of lensed galaxies via SED
fitting. Our conclusions can be summarized as follows:
\begin{itemize}
\item The background sources comprises a mixture of ULIRGs and HyLIRG
  with star formation rates SFR$\sim400-2000\,$M$_{\odot}$\,yr$^{-1}$, and large dust
  masses, $M_{\rm dust}=(7-30)\times10^{8}\,$M$_{\odot}$. SDP.11 and SDP.17 are resolved into multiple knots of rest-frame
  UV/optical emission in the source plane (D13) indicative of either
  a major merger \citep{Fu13} or distinct clumps of star formation within the same
  proto-galaxy \citep{Swin11}.
\item The lensed galaxies have already assembled a mass in stars
  $M_{\star}=(6-25)\times10^{10}\,$M$_{\odot}$. Their molecular gas content
  is still significant, $f_{\rm gas}\sim15-30\%$, so that
  star formation can be sustained for another $\lsim100\,$Myr  at the inferred rate. 
\item By the end of their star formation activity all these galaxies will
  have a mass in stars $\gsim10^{11}\,$M$_{\odot}$. We are thus witnessing the very
  early stages in the formation of elliptical galaxies, during the peak epoch ($z\sim1.5-3$) of the cosmic star formation history
  of the Universe. 
\end{itemize}
The wide range of lens-to-source flux
density ratios at 1.1- and 1.6-$\mu$m observed in this sample suggests that,  in some cases, the lensed source may significantly
contribute to the near-IR photometry of the system, as measured in low
angular resolution VIKING and  WISE surveys.
Therefore, sub-mm lens candidates showing an ``excess'' of emission at near-IR wavelengths compared
to what expected for a passively evolving elliptical (i.e. the lens) are ideal
targets for succesfull follow-up observations in the
near-IR with HST/WFC3 and with the Keck
telescopes (in Adaptive Optics), aiming at spatially resolving the
lensed structue in these systems \citep{Gonz12}.
Many more lens candidates from both H-ATLAS and HerMES have been now
observed with HST/WFC3/F110W in cycle-19 (PID: 12488)  and with Keck/AO in the $H$
and $K$ bands. Lens modeling and SED
fitting for these targets will be presented in a series of upcoming papers
(Amber et al. in prep.; Calanog et al. in prep.).

$~$ \\
{\bf Acknowledgments} \\
This work was supported by STFC (grants PP/D002400/1 and
ST/G002533/1), by ASI/INAF agreement I/072/09/0, by PRIN-INAF 2012
project ``Looking into the dust- obscured phase of galaxy formation
through cosmic zoom lenses in the Herschel Astrophysical Large Area
Survey'' and, in part, by the Spanish Ministerio de Ciencia e
Innovacion (project AYA2010-21766-C03- 01).
JGN acknowledges financial support from the Spanish CSIC for a JAE-DOC
fellowship, co-funded by the European Social Fund.
Herschel is an ESA space observatory with science instruments provided by European-led Principal Investigator consortia and with important participation from NASA.
The {\it Herschel}-ATLAS is a project with {\it Herschel}, which is an
ESA space observatory with science instruments provided by
European-led Principal Investigator consortia and with important
participation from NASA. The H-ATLAS website is
http://www.h-atlas.org/. 
This publication makes use of data products from the Wide-field Infrared Survey Explorer, which is a joint project of the University of California, Los Angeles, and the Jet Propulsion Laboratory/California Institute of Technology, funded by the National Aeronautics and Space Administration.

\label{lastpage}

\end{document}